\documentclass[aps,pra,reprint,superscriptaddress,longbibliography]{revtex4-1}
\pdfoutput=1
\usepackage{pdfpages}
\usepackage{graphicx}
\usepackage{morefloats}
\usepackage{color}
\usepackage{amssymb}

\makeatletter
\AtBeginDocument{\let\LS@rot\@undefined}
\makeatother

\begin{document}

\newcommand{\ham}{\mathcal{H}} 
\newcommand{\Q}{\mathcal{Q}} 

\title{Giant resonant nonlinear damping in nanoscale ferromagnets} 

\author{I.\,Barsukov}
\email[]{igorb@ucr.edu}
\affiliation{Physics and Astronomy, University of California, Irvine, CA 92697, USA}
\author{H. K. Lee}
\affiliation{Physics and Astronomy, University of California, Irvine, CA 92697, USA}
\author{A. A. Jara}
\affiliation{Physics and Astronomy, University of California, Irvine, CA 92697, USA}
\author{Y.-J.\,Chen}
\affiliation{Physics and Astronomy, University of California, Irvine, CA 92697, USA}
\author{A.\,M.\,Gon\c{c}alves}
\affiliation{Physics and Astronomy, University of California, Irvine, CA 92697, USA}
\author{C. Sha}
\affiliation{Physics and Astronomy, University of California, Irvine, CA 92697, USA}
\author{J.\,A.\,Katine}
\affiliation{Western Digital, 5600 Great Oaks Parkway, San Jose, CA 95119, USA}
\author{R.\,E.\,Arias}
\affiliation{Departamento de F\'{i}sica, CEDENNA, FCFM, Universidad de Chile, Santiago, Chile}
\author{B.\,A.\,Ivanov}
\affiliation{Institute of Magnetism, National Academy of Sciences of Ukraine, Vernadsky av. 36 B, Kyiv, 03142, Ukraine}
\affiliation{National University of Science and Technology MISiS, Moscow, 119049, Russian Federation}
\author{I.\,N.\,Krivorotov}
\affiliation{Physics and Astronomy, University of California, Irvine, CA 92697, USA}

\keywords{nonlinear damping, spin torque, nanomagnet, spin wave, ferromagnetic resonance}

\begin{abstract}
Magnetic damping is a key metric for emerging technologies based on magnetic nanoparticles, such as spin torque memory and high-resolution biomagnetic imaging. Despite its importance, understanding of magnetic dissipation in nanoscale ferromagnets remains elusive, and the damping is often treated as a phenomenological constant. Here we report the discovery of a giant frequency-dependent nonlinear damping that strongly alters the response of a nanoscale ferromagnet to spin torque and microwave magnetic field. This novel damping mechanism originates from three-magnon scattering that is strongly enhanced by geometric confinement of magnons in the nanomagnet. We show that the giant nonlinear damping can invert the effect of spin torque on a nanomagnet leading to a surprising current-induced enhancement of damping by an antidamping torque. Our work advances understanding of magnetic dynamics in nanoscale ferromagnets and spin torque devices.
\end{abstract}

\maketitle

\section{Introduction}
Nanoscale magnetic particles are the core components of several emerging technologies such as nonvolatile spin torque memory \cite{liu_spin-torque_2012}, spin torque oscillators \cite{kiselev2003microwave, rippard_spin-transfer_2010, houssameddine_spin-torque_2007,houshang_spin-wave-beam_2016, demidov_magnetic_2012, macia_stable_2014}, targeted drug delivery, and high-resolution biomagnetic imaging \cite{TsourkasScience, HoScience, Parak, Yilmaz}. Control of magnetic damping holds the key to improving the performance of many nanomagnet-based practical applications. In biomagnetic characterization techniques such as magnetic resonance imaging \cite{Shao2010}, relaxometry \cite{ConollyAPL}, and magnetic particle imaging \cite{Conolly, Krishnan}, magnetic damping affects nanoparticle’s relaxation times and image resolution. In spin torque memory and oscillators, magnetic damping determines the electrical current necessary for magnetic switching \cite{liu_spin-torque_2012} and generation of auto-oscillations \cite{slavin2009nonlinear} and thereby determines energy-efficiency of these technologies. The performance of nanomagnet-based microwave detectors is also directly affected by the damping \cite{zhu_voltage-induced_2012, miwa_highly_2014, fang_giant_2016}. Despite its importance across multiple disciplines, magnetic damping in nanoparticles is poorly understood and is usually modeled as a phenomenological constant \cite{slavin2009nonlinear,demidov_magnetic_2012}. 

In this article, we experimentally demonstrate that a ferromagnetic nanoparticle can exhibit dynamics qualitatively different from those predicted by the constant damping model. We show that nonlinear contributions to the damping can be unusually strong and the damping parameter itself can exhibit resonant frequency dependence. Our work demonstrates that nonlinear damping in nanomagnets is qualitatively different from that in bulk ferromagnets and requires a new theoretical framework for its description. We show both experimentally and theoretically that such resonant nonlinear damping originates from multi-magnon scattering in a magnetic system with a discrete spectrum of magnons induced by geometric confinement.

We also discover that the resonant nonlinear damping dramatically alters the response of a nanomagnet to spin torque. Spin torque arising from injection of spin currents polarized opposite to the direction of magnetization acts as negative damping \cite{kiselev2003microwave}. We find, however, that the effect of such antidamping spin torque is reversed, leading to an enhanced dissipation due to the nonlinear resonant scattering. This counterintuitive behavior should have significant impact on the operation of spin torque based memory \cite{liu_spin-torque_2012}, oscillators \cite{kiselev2003microwave, rippard_spin-transfer_2010, houssameddine_spin-torque_2007,houshang_spin-wave-beam_2016, demidov_magnetic_2012, macia_stable_2014} and microwave detectors \cite{zhu_voltage-induced_2012, miwa_highly_2014, fang_giant_2016}. 

\begin{figure*}[pt]
\includegraphics[width=0.99\textwidth]{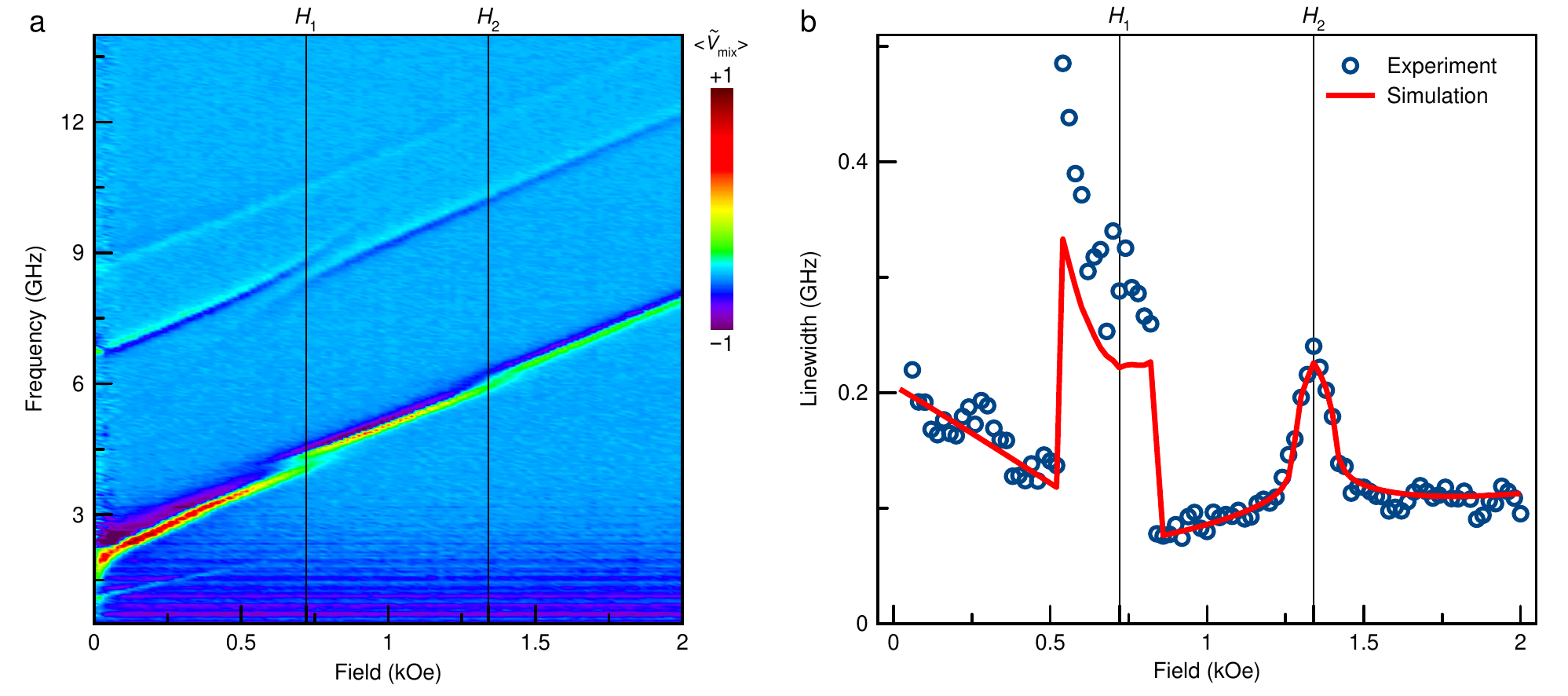}
\caption{Spin wave spectra in a nanoscale MTJ. (a)~Normalized ST-FMR spectra $\langle \tilde{V}_{\mathrm{mix}}(f) \rangle$ of spin wave eigenmodes in a perpendicular MTJ device (Sample~1) measured as a function of out-of-plane magnetic field. Resonance peaks arising from three low frequency modes of the MTJ free layer $|0\rangle$, $|1\rangle$, and $|2\rangle$ are observed. (b)~Spectral linewidth of the quasi-uniform $|0\rangle$ spin wave mode as a function of out-of-plane magnetic field. Strong linewidth enhancement is observed in the resonant three-magnon regime at $H_1$ and $H_2$.} 
\end{figure*} 

\section{Results}
\subsection{Spin wave spectroscopy}
We study nonlinear spin wave dynamics in nanoscale elliptical magnetic tunnel junctions (MTJs) that consist of a CoFeB free layer (FL), an MgO tunnel barrier, and a synthetic antiferromagnet (SAF) pinned layer \cite{AlexFMR}. Spectral properties of the FL spin wave modes are studied in a variety of MTJs with both in-plane and perpendicular-to-plane equilibrium orientations of the FL and SAF magnetization. We observe strong resonant nonlinear damping in both the in-plane and the perpendicular MTJs, which points to the universality of the effect. 

We employ spin torque ferromagnetic resonance (ST-FMR) to measure magnetic damping of the FL spin wave modes. In this technique, a microwave drive current $I_{\mathrm{ac}} \sin(2\pi f t)$ applied to the MTJ excites oscillations of magnetization at the drive frequency $f$. The resulting magnetoresistance oscillations $R_{\mathrm{ac}} \sin(2\pi f t+ \phi)$ generate a direct voltage $V_{\mathrm{mix}}$. Peaks in ST-FMR spectra $V_{\mathrm{mix}}(f)$ arise from resonant excitation of spin wave eigenmodes of the MTJ \cite{tulapurkar2005spin,Sankey,_damping_2016,safranski_material_2016,harder_electrical_2016,mosendz_quantifying_2010, liu_electrical_2014, MiwaPRX}. To improve signal-to-noise ratio, the magnitude of external magnetic field $H$ applied parallel to the free layer magnetization is modulated, and a field-derivative signal $\tilde{V}_{\mathrm{mix}}(f)=\mathrm{d} V_{\mathrm{mix}}(f)/\mathrm{d} H$ is measured via lock-in detection technique \cite{AlexFMR}. $V_{\mathrm{mix}}(f)$ can then be obtained via numerical integration (Supplemental Material). 

Figure~1(a) shows ST-FMR spectra $\tilde{V}_{\mathrm{mix}}(f)$ measured as a function of out-of-plane magnetic field $H$ for an elliptical 52\,nm\,$\times\,$62\,nm perpendicular MTJ device (Sample~1). Three spin wave eigenmodes with nearly linear frequency-field relation $f_n(H)$ are clearly visible in the spectra. Micromagnetic simulations (Supplemental Material) reveal that these modes are three lowest frequency spin wave eigenmodes of the FL (Supplemental Material). The lowest frequency (quasi-uniform) mode $|0\rangle$ is nodeless and has spatially uniform phase. Each of the two higher-order modes $|n\rangle$ ($n=1,2$) has a single node at the FL center that is either perpendicular ($n=1$) or parallel ($n=2$) to the ellipse long axis. 

The spectral linewidth of the resonances in Fig.\,1(a) can be used for evaluation of the mode damping. The quasi-uniform mode $|0\rangle$ resonance visibly broadens at two magnetic field values: $H_{1}=0.74$\,kOe (4\,GHz) and $H_{2}=1.34$\,kOe (6\,GHz). Near $H_1$, the mode $|1\rangle$ resonance also broadens and exhibits splitting, same behavior is observed for the mode $|2\rangle$ at $H_2$. At these fields, the higher-order mode frequency is twice that of the quasi-uniform mode $f_n=2f_0$. This shows that three-magnon confluence \cite{Patton3m,schultheiss_direct_2009,Boone,kurebayashi3m,
CostaFilho} is the mechanism of the quasi-uniform mode damping increase: two magnons of the quasi-uniform mode $|0\rangle$ merge into a single magnon of the higher-order mode $|n\rangle$.

\begin{figure*}[pt]
\includegraphics[width=1.0\textwidth]{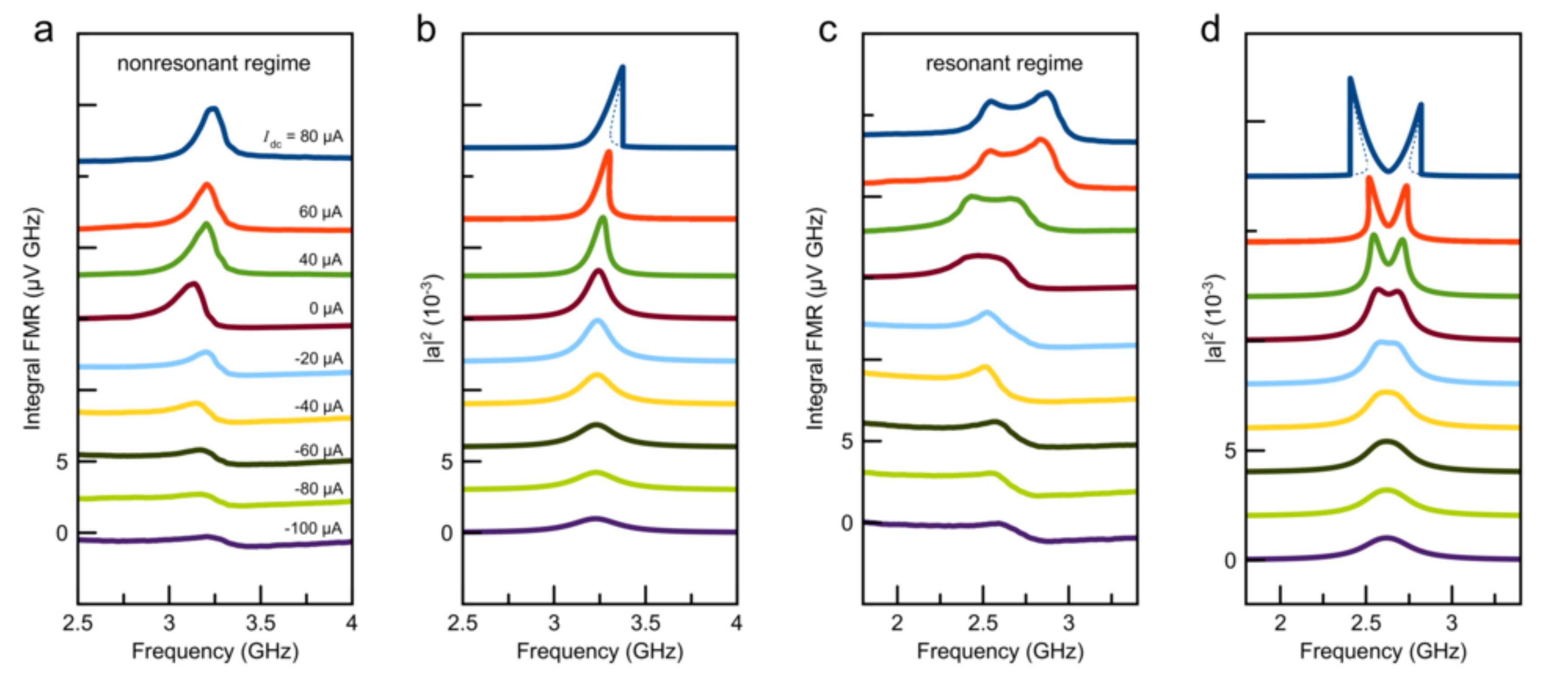}
\caption{Effect of spin torque on spin wave resonance lineshape. (a)-(b)~Spin wave resonance lineshapes in the nonresonant regime at $H> H_1$ for different values of direct bias current $I_\mathrm{dc}$. (c)-(d)~Spin wave resonance lineshapes in the resonant three-magnon regime at $H = H_1$. (a), (c)~Measured ST-FMR spectra (Sample 2). (b), (d)~Solutions of Eqs.~(3) and (4). Identical bias current values $I_\mathrm{dc}$ (displayed in (a) are used in (a)-(d).}
\end{figure*} 

The most striking feature of the quasi-uniform mode resonance near $H_1$ is its split-peak shape with a local minimum at the resonance frequency. Such a lineshape cannot be fit by the standard Lorentzian curve with symmetric and antisymmetric components \cite{AlexFMR}. We therefore use a double-peak fitting function (Supplemental Material) to quantify the effective linewidth $\Delta f_0$ of the resonance profile. For applied fields sufficiently far from $H_1$, the ST-FMR curve recovers its single-peak shape and $\Delta f_0$ is determined as half width of the standard Lorentzian fitting function \cite{AlexFMR}. Figure~1(b) shows $\Delta f_0$ as a function of $H$ and demonstrates a large increase of the linewidth near the fields of the resonant three-magnon regime $H_1$ and $H_2$. The stepwise increase of $\Delta f_0$ near $H_1$ is a result of the ST-FMR curve transition between the split-peak and single-peak shapes. For fields near $H_2$, the resonance profile broadens but does not develop a visible split-peak lineshape. As a result, $\Delta f_0(H)$ is a smooth function in the vicinity of $H_2$. 

\subsection{Effect of spin torque} In MTJs, direct bias current $I_\mathrm{dc}$ applied across the junction exerts spin torque on the FL magnetization, acting as antidamping for $I_\mathrm{dc}>0$ and as positive damping for $I_\mathrm{dc}<0$ \cite{Sankey,deac}. 
The antidamping spin torque increases the amplitude of the FL spin wave modes \cite{Sankey,demidov_control_2011} and decreases their spectral linewidth \cite{fuchs_spin-torque_2007}. We can employ spin torque from $I_\mathrm{dc}$ to control the amplitude of spin wave eigenmodes excited in ST-FMR measurements, and thereby study the crossover between linear and nonlinear regimes of spin wave resonance. 

Figure\,2 shows the dependence of ST-FMR resonance curve of the $|0\rangle$ mode $V_{\mathrm{mix}}(f)$ on $I_\mathrm{dc}$ for a 50\,nm$\,\times\,$110\,nm elliptical in-plane MTJ (Sample 2). For in-plane magnetic field values far from the three-magnon resonance fields $H_{n}$, the amplitude of ST-FMR resonance curve $V_{\mathrm{mix}}(f)$ shown in Fig.\,2(a) monotonically increases with increasing antidamping spin torque, as expected. At $H = H_{1}$, the antidamping spin torque has a radically different and rather surprising effect on the resonance curve. As illustrated in Fig.\,2(c), increasing antidamping spin torque first broadens the resonance at $H = H_{1}$ and then transforms a single-peak resonance lineshape into a split-peak lineshape with a local minimum at the resonance frequency $f_0$.  The data in Fig.\,2 demonstrate that the unusual split-peak lineshape of the resonance is only observed when (i)~the three-magnon scattering of the quasi-uniform mode is allowed by the conservation of energy and (ii)~the amplitude of the mode is sufficiently high, confirming that the observed effect is resonant and nonlinear in nature.

Fig.\,2(c) reveals that antidamping spin torque can increase the spectral linewidth and the effective damping of the quasi-uniform spin mode if the mode undergoes resonant three-magnon scattering. Figure~3 further illustrates this counterintuitive effect. It shows the linewidth of the quasi-uniform mode of a 50\,nm$\,\times$110\,nm elliptical in-plane MTJ (Sample~3) measured as a function of bias current. In Fig.\,3, blue symbols show the linewidth measured at an in-plane magnetic field sufficiently far from the three-magnon resonance fields $H_n$. At this field, the expected quasi-linear dependence of the linewidth on $I_{\mathrm{dc}}$ is observed for currents well below the critical current for the excitation of auto-oscillatory magnetic dynamics. Near the critical current, the linewidth increases due to a combination of the fold-over effect \cite{melkov2013nonlinear,helsen_non-linear_2015, podbielski_microwave-assisted_2007} and thermally activated switching between the large- and small-amplitude oscillatory states of the fold-over regime \cite{Sankey}. The red symbols in Fig.\,3  show the linewidth measured in the resonant three-magnon regime at $H=H_1$. In contrast to the nonresonant regime, the linewidth increases with increasing $|I_\mathrm{dc}|$ for both current polarities. Furthermore, the maximum linewidth is measured for the antidamping current polarity.

\section{Theoretical model}
Nonlinear interactions among spin wave eigenmodes of a ferromagnet give rise to a number of spectacular magneto-dynamic phenomena such as Suhl instability of the uniform precession of magnetization \cite{SUHL1957209, bauer_nonlinear_2015}, spin wave self-focusing \cite{SlavinFocusing} and magnetic soliton formation \cite{ Kosevich1990,SlavinSoliton, wu_self-generation_2005}. In bulk ferromagnets, nonlinear interactions generally couple each spin wave eigenmode to a continuum of other modes via energy- and momentum-conserving multi-magnon scattering \cite{SUHL1957209}. This kinematically allowed scattering limits the achievable amplitude of spin wave modes and leads to broadening of the spin wave resonance. These processes lead to a resonance broadening \cite{SUHL1957209,NaletovNonlin,khivintsev_nonlinear_2010, rana_effect_2017} and cannot explain the observed split-peak lineshape of the resonance. In nanoscale ferromagnets, geometric confinement discretizes the spin wave spectrum and thereby generally eliminates the kinematically allowed multi-magnon scattering.  This suppression of nonlinear scattering enables persistent excitation of spin waves with very large amplitudes \cite{daquino_analytical_2017} as observed in nanomagnet-based spin torque oscillators \cite{kiselev2003microwave, krivorotov_time-domain_2008}. Tunability of the spin wave spectrum by external magnetic field, however, can lead to a resonant restoration of the energy-conserving scattering \cite{Boone}. The description of nonlinear spin wave resonance in the nanoscale ferromagnet geometry therefore requires a new theoretical framework. To derive the theory of resonant nonlinear damping in a nanomagnet, we start with a model Hamiltonian that explicitly takes into account resonant nonlinear scattering between the quasi-uniform mode and a higher-order spin wave mode (in reduced units with $\hbar \equiv1$):
\begin{eqnarray}
\label{Eq1}
\ham = \omega_0 a^\dagger a + \omega_n b^\dagger b 
+\frac{\Psi_0}{2} a^\dagger a^\dagger a a +\frac{\Psi_n}{2} b^\dagger b^\dagger b b \\ \nonumber
+(\psi_{n} a a b^\dagger +\psi_{n}^* a^\dagger a^\dagger b) \\ \nonumber
+ \zeta\left\{\exp(-i\,\omega  t)a^\dagger+\exp(i\,\omega t)a\right\} 
\end{eqnarray}
where $a^\dagger$, $a$ and $b^\dagger$, $b$ are the magnon creation and annihilation operators for the quasi-uniform mode $|0\rangle$ with frequency $\omega_0$ and for the higher-order spin wave mode $\left|n\right\rangle$ mode with frequency $\omega_n$, respectively. The nonlinear mode coupling term proportional to the coupling strength parameter $\psi_n$ describes the annihilation of two $\left|0\right\rangle$ magnons and creation of one $\left|n\right\rangle$ magnon, as well as the inverse process. The Hamiltonian is written in the resonant approximation, where small nonresonant terms such as $aab$, $aaa^\dagger$ are neglected. The terms proportional to $\Psi_0$ and $\Psi_n$ describe the intrinsic nonlinear frequency shifts \cite{Guo} of the modes $|0\rangle$ and $|n\rangle$. The last term describes the excitation of the quasi-uniform mode by an external ac drive with the amplitude $\zeta$ and frequency $\omega$.

We further define classically a dissipation function $\Q$, where $\alpha_0$ and $\alpha_n$ are the intrinsic linear damping parameters of the modes $|0\rangle$ and $|n\rangle$ \cite{nembach_mode-_2013, li_wave-number-dependent_2016, sekiguchi_time-domain_2012}: 
\begin{eqnarray}
\Q =  \frac{ \mathrm{d} a^\dagger}{\mathrm{d} t} \frac{ \mathrm{d} a}{\mathrm{d} t}(\alpha_0+\eta_0 a^\dagger a) +  \frac{ \mathrm{d} b^\dagger}{\mathrm{d} t} \frac{ \mathrm{d} b}{\mathrm{d} t} (\alpha_n+\eta_n b^\dagger b)
\end{eqnarray}

For generality, Eq.\,(2) includes intrinsic nonlinear damping \cite{slavin2009nonlinear} of the modes $|0\rangle$ and $|n\rangle$ described by the nonlinearity parameters $\eta_0$ and $\eta_n$. However, our analysis below shows that the split-peak resonance lineshape is predicted by our theory even if $\eta_0$ and $\eta_n$ are set equal to zero.

Equations describing the nonlinear dynamics of the two coupled spin wave modes of the system follow from Eq.~(1) and Eq.~(2):
\begin{eqnarray}
i \frac{ \mathrm{d}a}{\mathrm{d} t} = \frac{\partial \ham}{\partial a^\dagger} + \frac{\partial \Q}{ \partial (\mathrm{d} a^\dagger /\mathrm{d} t)} \\ 
i \frac{ \mathrm{d} b}{\mathrm{d} t} = \frac{\partial \ham}{\partial b^\dagger} + \frac{\partial \Q}{ \partial (\mathrm{d} b^\dagger /\mathrm{d} t)}
\end{eqnarray}  

\begin{figure}[pt]
\includegraphics[width=0.99\columnwidth]{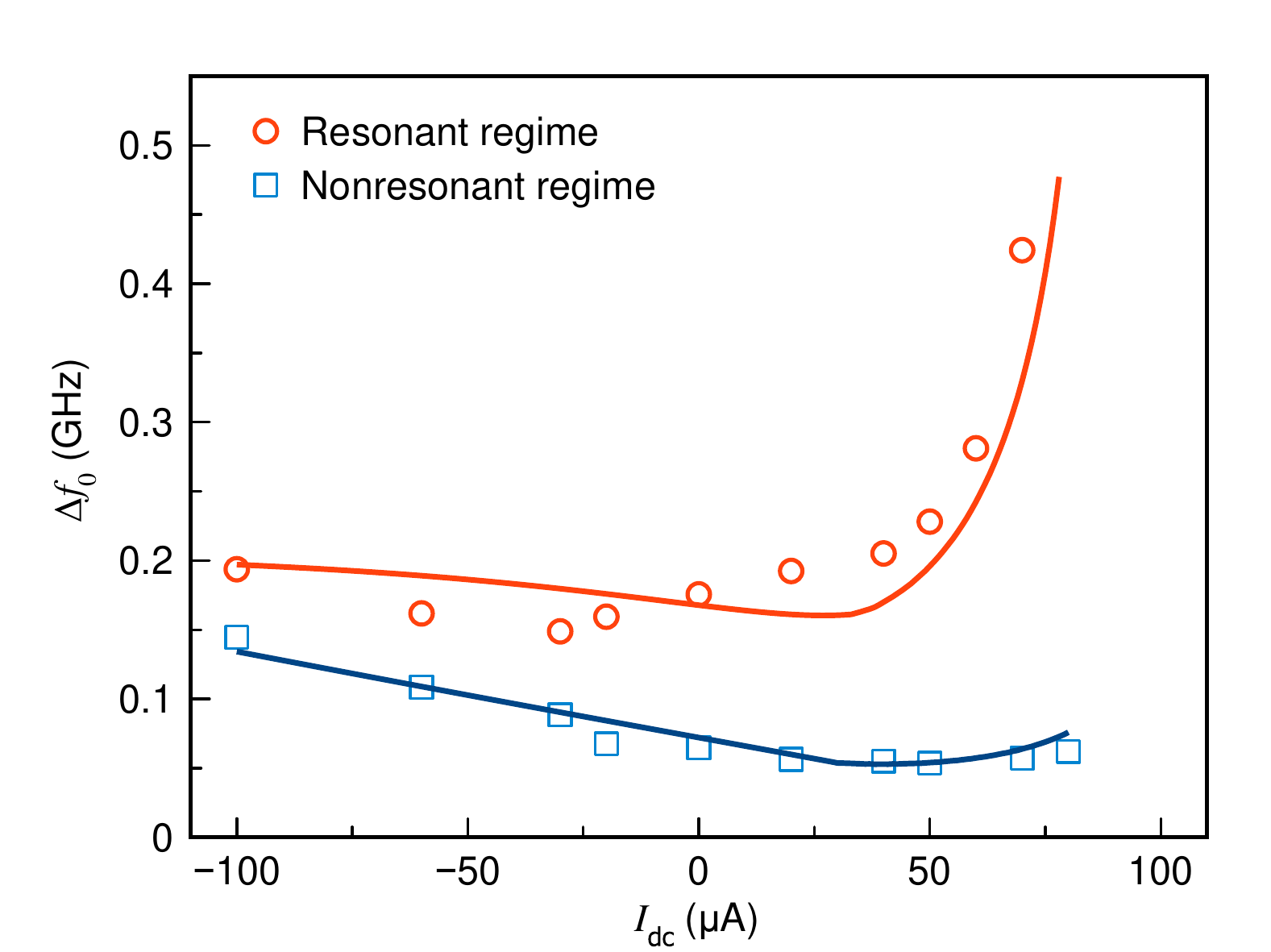}
\caption{Effect of spin torque on linewidth. Linewidth of the quasi-uniform spin wave mode as a function of the applied direct bias current (Sample~3): blue symbols -- in the nonresonant regime $H\neq H_{1}$ and red symbols -- in the resonant three-magnon regime $H= H_{1}$. Lines are numerical fits using Eqs.~(3) and (4).}
\end{figure} 

It can be shown (Supplemental Material) that these equations have a periodic solution  $a = \bar a \exp{(-i\,\omega t)}$ and  $b = \bar b \exp{(-i\,2\omega t)}$, where $\bar a$, $\bar b$ are the complex spin wave mode amplitudes. For such periodic solution, Eqs.~(3) and (4) are reduced to a set of two nonlinear algebraic equations for absolute values of the spin wave mode amplitudes $|\bar{a}|$ and $|\bar{b}|$, which can be solved numerically. Since the ST-FMR signal is proportional to $|\bar{a}|^2$ (Supplemental Material), the calculated $|\bar{a}|^2(\omega)$ function can be directly compared to the measured ST-FMR resonance lineshape.

We employ the solution of Eqs.~(3) and (4) to fit the field dependence of the quasi-uniform mode linewidth in Fig.\,1(b). In this fitting procedure, the resonance lineshape $|\bar{a}|^2(\omega)$ is calculated, and its spectral linewidth $\Delta \omega_0$ is found numerically. The resonance frequencies $\omega_0$ and $\omega_n$ are directly determined from the ST-FMR data in Fig.\,1(a). The intrinsic damping parameters $\alpha_0$ and $\alpha_n$ near $H_1$ and $H_2$ are found from linear interpolations of the ST-FMR linewidths $\Delta f_0$ and $\Delta f_n$ measured at fields far from $H_1$ and $H_2$.  We find that $\Delta \omega_0$ weakly depends on the nonlinearity parameters $\Psi$ and $\eta$, and thus these parameters are set to zero (Supplemental Material). We also find that the calculated linewidth $\Delta \omega_0$  depends on the product of the drive amplitude $\zeta $ and mode coupling strength $\psi_n$, but is nearly insensitive to the individual values of $\zeta$ and $\psi_n$ as long as $\zeta \cdot \psi_n$ = const (Supplemental Material). Therefore, we use $\zeta \cdot \psi_n$ as a single fitting parameter in this fitting procedure. Solid line in Fig.\,1(b) shows the calculated field dependence of the quasi-uniform mode linewidth on magnetic field. The agreement of this single-parameter fit with the experiment is excellent.

Figures 2(b) and 2(d) illustrate that Eqs.~(3) and (4) not only describe the field dependence of ST-FMR linewidth but also qualitatively reproduce the spectral lineshapes of the measured ST-FMR resonances as well as the effect of the antidamping spin torque on the lineshapes. Fig.\,2(b) shows the dependence of the calculated lineshape $|\bar{a}|^2(\omega)$ on antidamping spin torque for a magnetic field $H$ far from the three-magnon resonance fields $H_n$. At this nonresonant field, increasing antidamping spin torque induces the fold-over of the resonance curve \cite{melkov2013nonlinear} without resonance peak splitting. The dependence of $|\bar{a}|^2(\omega)$ on antidamping spin torque for $H=H_1$ is shown in Fig.\,2(d). At this field, the resonance peak in $|\bar{a}|^2(\omega)$ first broadens with increasing antidamping spin torque and then splits, in qualitative agreement with the experimental ST-FMR data in Fig.\,2(c). Our calculations (Supplemental Material) reveal that while the nonlinearity parameters $\Psi_0$, $\eta_0$, $\Psi_n$ and $\eta_n$ have little effect on the linewidth $\Delta \omega_0$, they modify the lineshape of the resonance. Given that the nonlinearity parameter values are not well known for the systems studied here, we do not attempt to quantitatively fit the measured ST-FMR lineshapes. 

Equations~(3) and (4) also quantitatively explain the observed dependence of the quasi-uniform mode linewidth $\Delta \omega_0$ on direct bias current $I_\mathrm{dc}$. Assuming antidamping spin torque linear in bias current \cite{fuchs_spin-torque_2007,lauer_spin-transfer_2016,zhang_engineering_2017}: $\alpha_0 \rightarrow \alpha_0(1-I_\mathrm{dc}/I^{|0\rangle}_\mathrm{c})$,  $\alpha_n \rightarrow \alpha_n(1-I_\mathrm{dc}/I^{|n\rangle}_\mathrm{c})$, where $I^{|n\rangle}_\mathrm{c}>I^{|0\rangle}_\mathrm{c}$ are the critical currents, we fit the measured bias dependence of ST-FMR linewidth in Fig.~3 by solving Eqs.~(3) and (4). The solid lines in Fig.~3 are the best numerical fits, where $\zeta \cdot \psi_n$ and $I_\mathrm{c}$ are used as independent fitting parameters. The rest of the parameters in Eqs.~(3) and (4) are directly determined from the experiment following the procedure used for fitting the data in Fig.~1(b). Theoretical curves in Fig.~3 capture the main feature of the data at the three-magnon resonance field $H_1$ -- increase of the linewidth with increasing antidamping spin torque.

\section{Discussion}
Further insight into the mechanisms of the nonlinear spin wave resonance peak splitting and broadening by antidamping spin torque can be gained by neglecting the intrinsic nonlinearities $\Psi_n$ and $\eta_n$ of the higher-order mode $|n\rangle$. 
Setting $\Psi_n=0$ and $\eta_n=0$ in Eqs.~(3) and (4) allows us to reduce the equation of motion for the quasi-uniform mode amplitude $|\bar{a}|$ to the standard equation for a single-mode damped driven oscillator (Supplemental Material) where a constant damping parameter $\alpha_0$ is replaced by an effective frequency-dependent nonlinear damping parameter $\alpha_0^{\mathrm{eff}}$:
\begin{equation}
\alpha_0^{\mathrm{eff}} = \alpha_0+ \left[ \eta_0 +\frac{4\alpha_n  \psi_n^2}{(2\omega-\omega_n)^2+4\alpha_n ^2\omega ^2} \right] |\bar a|^2
\end{equation}
and the resonance frequency is replaced by an effective resonance frequency:
\begin{equation}
\omega_0^{\mathrm{eff}} =  \omega_0 + \left[ \Psi_0 + \frac{2|\psi_n|^2 (2\omega-\omega_n)}{(2\omega-\omega_n)^2 + 4\alpha_n^2\omega^2} \right] |\bar{a}|^2
\end{equation}

Equation~(5) clearly shows that the damping parameter of the quasi-uniform mode itself becomes a resonant function of the drive frequency with a maximum at half the frequency of the higher order mode ($\omega=\frac{1}{2}\omega_n$). The amplitude and the width of this resonance in $\alpha_0^{\mathrm{eff}}(\omega)$ are determined by the intrinsic damping parameter $\alpha_n$ of the higher-order mode $|n\rangle$. If $\alpha_n$ is sufficiently small, the quasi-uniform mode damping is strongly enhanced at $\omega=\frac{1}{2}\omega_n$, which leads to a decrease of the quasi-uniform mode amplitude at this drive frequency. If the drive frequency is shifted away from $\frac{1}{2}\omega_n$ to either higher or lower values, the damping decreases, which can result in an increase of the quasi-uniform mode amplitude $|\bar{a}|$. Therefore, the amplitude of the quasi-uniform mode $|\bar{a}|(\omega)$ can exhibit a local minimum at $\omega=\frac{1}{2}\omega_n$. Due to its nonlinear origin, the tendency to form a local minimum in $|\bar{a}|(\omega)$ at $\frac{1}{2}\omega_n$ is enhanced with increasing $|\bar a|$. Since $|\bar a|$ is large near the resonance frequency $\omega_0$, tuning $\omega_0$ to be equal to $\frac{1}{2}\omega_n$ greatly amplifies the effect of local minimum formation in $|\bar{a}|(\omega)$. This qualitative argument based on Equation (5) explains the data in Fig.~2 -- the split-peak nonlinear resonance of the quasi-uniform mode is only observed when external magnetic field tunes the spin wave eigenmode frequencies to the three-magnon resonance condition $\omega_0=\frac{1}{2}\omega_n$.

Equation~(6) reveals that the nonlinear frequency shift of the quasi-uniform mode is also a resonant function of the drive frequency. In contrast to the nonlinear damping resonance described by Equation (5), the frequency shift resonance is an antisymmetric function of $\omega-\frac{1}{2}\omega_n$. The nonlinear shift is negative for $\omega<\frac{1}{2}\omega_n$ and thus causes a fold-over towards lower frequencies while it is  positive for $\omega>\frac{1}{2}\omega_n$ causing fold-over towards higher frequencies. At the center of the resonance profile, the three-magnon process induces no frequency shift. This double-sided fold-over also contributes to the formation of the split-peak lineshape of the resonance shown in Figs.~2(c) and 2(d) and to the linewidth broadening. As with the nonlinear damping resonance, the antisymmetric nonlinear frequency shift and the double-sided fold-over become greatly amplified when the spin wave mode frequencies are tuned near the three-magnon resonance $\omega_0=\frac{1}{2}\omega_n$. 

Equations (5) and (6) also shed light on the origin of the quasi-uniform mode line broadening by the antidamping spin torque. The antidamping spin torque increases the quasi-uniform mode amplitude $|\bar a|$ via transfer of angular momentum from spin current to the mode \cite{rezende_magnon_2006}. Since the nonlinear damping and the nonlinear frequency shift are both proportional to $|\bar a|^2$ and both contribute to the line broadening, the antidamping spin torque can indeed give rise to the line broadening. Equation (5) reveals two competing effects of the antidamping spin torque on the quasi-uniform mode damping parameter $\alpha_0^{\mathrm{eff}}$: spin torque from $I_\mathrm{dc}$ decreases the linear component of the damping parameter $\alpha_0 \rightarrow \alpha_0(1-I_\mathrm{dc}/I^{|0\rangle}_\mathrm{c})$ and increases the nonlinear component via increased $|\bar a|^2$. Whether the antidamping spin torque decreases or increases the spectral linewidth of the mode depends on the system parameters. Our numerical solution of Eqs.~(3) and (4) shown in Fig.~3 clearly demonstrates that the antidamping spin torque can strongly increase the linewidth of the quasi-uniform mode when the three-magnon resonance condition $\omega_0=\frac{1}{2}\omega_n$ is satisfied. Furthermore, we find that the three-magnon process exhibits no threshold behavior upon increasing amplitude (Supplemental Material) or decreasing intrinsic damping. 

The key requirement for observation of the resonant nonlinear damping is the discreteness of the magnon spectrum imposed by geometric confinement in the nanoscale ferromagnet. The split-peak nonlinear resonance discovered in this work cannot be realized in bulk ferromagnets because the three-magnon resonance condition in bulk is not only valid at the uniform mode frequency $\omega_0=\frac{1}{2}\omega_n$ but instead in a broad frequency range. Owing to the magnon spectrum continuity in bulk, shifting the excitation frequency away from $\omega_0$ does not suppress the three-magnon scattering of the uniform mode -- it simply shifts it from one group of magnons to another \cite{SUHL1957209,Patton3m}. Therefore, the amplitude of the uniform mode does not increase when the drive frequency is shifted away from $\omega_0$ and the split-peak resonance is not realized.

We expect that the resonant nonlinear damping discovered in this work will have strong impact on the performance of spin torque devices such as spin torque magnetic memory, spin torque nanooscillators and spin torque microwave detectors. Since all these devices rely on large-amplitude oscillations of magnetization driven by spin torque, the amplitude limiting resulting from the resonant nonlinear damping is expected to have detrimental effect on the device performance. 

\section{Conclusions}
In conclusion, our measurements demonstrate that magnetic damping of spin wave modes in a nanoscale ferromagnet has a strong nonlinear component of resonant character that appears at a discrete set of magnetic fields corresponding to resonant three-magnon scattering. This strong resonant nonlinearity can give rise to unusual spin wave resonance profile with a local minimum at the resonance frequency in sharp contrast to the properties of the linear and nonlinear spin wave resonances in bulk ferromagnets. The resonant nonlinearity has a profound effect on the response of the nanomagnet to spin torque. Antidamping spin torque, that reduces the quasi-uniform spin wave mode damping at magnetic fields far from the resonant  three-magnon regime, can strongly enhance the damping in the resonant regime. This inversion of the effect of spin torque on magnetization dynamics by the resonant nonlinearity is expected to have significant impact on the performance of nanoscale spin torque devices such as magnetic memory and spin torque oscillators.

\begin{acknowledgments} 
This work was supported by the National Science Foundation through Grants No. DMR-1610146, No. EFMA-1641989 and No. ECCS-1708885. We also acknowledge support by the Army Research Office through Grant No. W911NF-16-1-0472 and Defense Threat Reduction Agency through Grant No. HDTRA1-16-1-0025. A.\,M.\,G. thanks CAPES Foundation, Ministry of Education of Brazil for financial support. R.E.A acknowledges Financiamiento Basal para Centros Cientificos y Tecnologicos de Excelencia under project FB 0807 (Chile), and Grant ICM P10-061-F by Fondo de Innovacion para la Competitividad-MINECON. B.A.I. was supported by the National Academy of Sciences of Ukraine via project \#1/17-N and by the Program of NUST "MISiS" (grant No.\,K2-2017-005), implemented by a governmental decree dated 16th of March 2013, No. 211.
\end{acknowledgments}

\newpage
\includepdf[pages={1}]{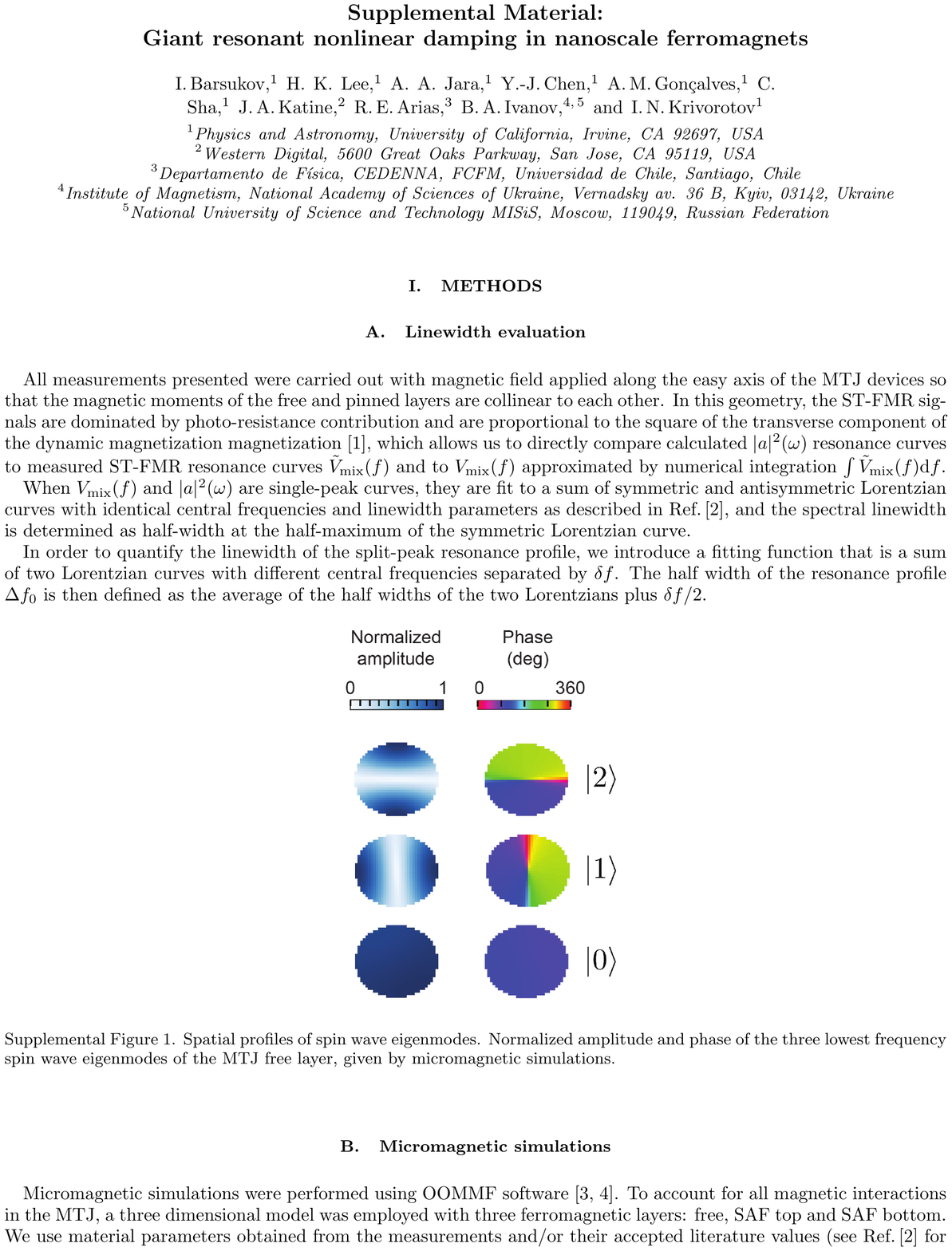}
~
\newpage
\includepdf[pages={2}]{supplemental}
~
\newpage
\includepdf[pages={3}]{supplemental}
~
\newpage
\includepdf[pages={4}]{supplemental}
~
\newpage
\includepdf[pages={5}]{supplemental}
~
\newpage
\includepdf[pages={6}]{supplemental}

\end{document}